\documentclass[aps,twocolumn,showpacs,floatfix,prl]{revtex4-1}
\usepackage{hyperref}
\hypersetup{colorlinks=true, urlcolor=blue}

\usepackage{epsfig}
\usepackage{newlfont}
\usepackage{amssymb}
\usepackage{amsfonts}
\usepackage{amsmath}
\usepackage{bm}

\newcommand{\Tr}{\mbox{tr}}

\def\lsim{\mathrel{\rlap{\lower4pt\hbox{\hskip1pt$\sim$}}
    \raise1pt\hbox{$<$}}}                % less than or approx. symbol
\def\gsim{\mathrel{\rlap{\lower4pt\hbox{\hskip1pt$\sim$}}
    \raise1pt\hbox{$>$}}}                % greater than or approx. symbol

\begin{document}

\renewcommand*{\DefineNamedColor}[4]{%
   \textcolor[named]{#2}{\rule{7mm}{7mm}}\quad
  \texttt{#2}\strut\\}

\definecolor{red}{rgb}{1,0,0}
\definecolor{cyan}{cmyk}{1,0,0,0}

%\title{Genuine multisite entanglement in large size isotropic multiqubit superpositions}
%\title{Density Matrix Recursion Method: \\ Characterizing genuine multisite entanglement in square spin lattices}
%\title{Characterizing Genuine Multisite Entanglement in Isotropic Square Spin Lattices}
\title{Characterizing Genuine Multisite Entanglement in Isotropic Spin Lattices}

\author{Himadri Shekhar Dhar\(^{1}\), Aditi Sen(De)\(^{2}\), and Ujjwal Sen\(^{2}\)}

\affiliation{\(^1\)School of Physical Sciences, Jawaharlal Nehru University, New Delhi 110 067, India\\
\(^2\)Harish-Chandra Research Institute, Chhatnag Road, Jhunsi, Allahabad 211 019, India
}

%\date{16 March 2011}

\begin{abstract}

We consider a class of large superposed states, obtained from dimer coverings on spin-1/2 isotropic lattices, whose potential usefulness ranges from organic molecules to quantum computation.
We show that they are genuinely multiparty entangled, irrespective of the geometry and dimension of the isotropic lattice.
We then present  an efficient method to characterize the genuine multisite entanglement in the case of isotropic square spin-1/2 lattices, with short-range dimer coverings. 
We use this iterative analytical method to calculate the multisite entanglement of finite-sized lattices, which, through finite-size scaling, enables us to obtain the estimate of the multisite entanglement of the infinite square lattice. 
The method can be a useful tool to investigate other single- and multisite properties of such states.
 
% containing relatively large number of spin sites. Such spin-1/2 rotationally invariant lattices can be shown to be always multisite globally entangled making them an important tool in many body quantum computation. 
% We use an iterative technique to generate the reduced density matrices for these large size spin systems and use a computable geometric measure called \emph{generalized geometric measure} to obtain the genuine multi-site entanglement. 
% We show that the characterization enables us to effectively calculate the multisite entanglement and other global correlation properties in arbitrary sized square quantum spin systems with short-range dimer coverings. The characterization 
% can be extended to other similar isotropic quantum spin states that contain large number of superposed dimer coverings.

\end{abstract}

%\pacs{03.67.-a, 75.10.Jm, 75.10.Pq, 05.30.Rt}

\maketitle 

%\section{Introduction}
%\label{intro}

\noindent\emph{Introduction.--} Genuine multisite entanglement is an important resource in quantum information protocols  and is known to offer significant advantage in quantum tasks in comparison to bipartite entanglement \cite{horo}.
In particular, it is the basic ingredient in measurement-based quantum computation \cite{computation}, and 
is beneficial
in various quantum communication protocols \cite{communicate}, including  secret sharing \cite{crypto} (cf.\,\cite{qss}). 
Apart from the conventional information tasks, the study of multisite entanglement turns out to be important in understanding many-body phenomena such as
quantum phase transitions \cite{QPT} and 
%in biological mechanisms 
to understand transport properties in the evolution of photosynthetic complexes \cite{bio2}.
Although bipartite entanglement in the case of two spin-1/2 particles is rather well-understood, the situation is quite different in the case of 
classification and quantification of entanglement in higher dimensions as well as in multiparty systems.
The fact that many-particle systems can have different 
types of useful entanglement, depending on the particular information processing protocols under study, makes the quantification a formidable task. 
%The number of quantifiers needed to compute entanglement in many body systems increases with system size \cite{MS3}. 
% are still limited. 
While there are several %\textcolor{red}
{characterizations of}
%However, there are several 
multiparty entanglement measures known in the literature \cite{horo, MS1}, it is in general difficult to compute %\textcolor{red}
{them for large systems.} However, for pure multisite states, it is possible to use the %\textcolor{red}
{\emph{generalized geometric 
measure} \cite{GGM}, which is a computable geometric measure of genuine multisite entanglement}. 

The isotropic spin-1/2 lattice states, with  short-range dimer coverings %\textcolor{red}
{are possible ground states of quantum spin liquids, known as resonating valence bond (RVB) states \cite{ander}}, and are of considerable interest in 
%investigating  organic molecules \cite{organic}, 
high-temperature superconductivity \cite{super}, 
%entanglement properties \cite{EP, asen_PRL} and 
cooperative phenomena in 
%\cite{critical} using information theory in condensed matter and 
many body systems \cite{critical, scal}, and quantum computation \cite{qc}. 
%However, there is no method that can suitably quantify the multisite entanglement in the square spin-1/2 lattices.
These short-range dimer states can be efficiently simulated in laboratories using atoms in optical lattices \cite{opt} or in cavities using interacting photons \cite{photon}. 
The large number of superpositions of the nearest-neighbor (NN) dimer coverings even in moderately large two-dimensional (2D) lattices, 
%for relatively large number of spins 
makes the %\textcolor{red}
{characterization and computation} of both bipartite and multisite entanglement of such spin states an arduous task.  

In this Letter, we consider quantum states, with
%short-range 
dimer coverings, of an isotropic spin-1/2 lattice.
% with a semi-periodic boundary condition. 
We prove that such spin-1/2 rotationally invariant states are always genuinely multisite 
%globally 
entangled, irrespective of the lattice geometry and dimension. 
We present a method to analytically calculate the genuine multipartite entanglement, viz. the generalized geometric measure \cite{GGM}, of these quantum spin states, with short-ranged dimer coverings, 
on a square lattice with an arbitrary number of sites.
The method enables us to 
%We use the method to 
calculate the genuine multisite entanglement for moderately large lattices, and perform finite-size scaling \cite{book-scale} to predict the measure for the infinite two-dimensional square lattice.
%
%The paper is organized as follows. The succeeding section provides a formal definition of the spin state that we consider. Section \ref{ME} contains the theorem proving that the spin state is 
%genuinely multiparty entangled. In section \ref{5:13}, we present a method to evaluate properties of such states on two-dimensional square lattices. In particular, a finite-size scaling is presented in Section \ref{fss}. We conclude with some discussions 
%in Section \ref{disc}.
%

An isotropic spin-1/2 system, with NN dimer coverings, 
%is 
%conveniently defined by  using a \emph{bipartite lattice}. 
%A bipartite lattice is one which is divided into two sublattices, 
%A and B, such that each site
%of sublattice A has only sites of sublattice B as its nearest neighbors. Moreover, each lattice site is occupied by a spin-1/2 particle. 
%% contains a spin-1/2 qubit and each qubit in sublattice A (B) can form a dimer with another qubit in sublattice B(A). 
%The dimer-covering state on such a bipartite lattice 
can be 
%The spin-1/2 state of $2N$ spin system can then be 
written in the form \cite{anders},
\begin{equation}
|\psi\rangle=\sum_k h_k(i_a,i_b)|(a_1,b_1),(a_2,b_2)...(a_N,b_N)\rangle_k,
\label{RVB}
\end{equation}
where each $k$ refers to a single dimer covering ($\{|a_i,b_i\rangle\}_i$) on the entire lattice
%between sites of the two sublattices, 
with $N$ spins in each sublattice of a bipartite lattice. 
Here, the \emph{covering function} $h_k(i_a,i_b)$ is isotropic over the lattice. For a detailed description of the bipartite spin lattice, see the Supplemental Material \cite{supple}.
Such states on a bipartite spin-1/2 lattice are known as RVB liquid states. In isotropic 2D systems, 
RVB-like dimer states are found to be ground states of 
a frustrated antiferromagnet on the 1/5-depleted square lattice
%CaV$_{\mathrm{4}}$O$_{\mathrm{9}}$ lattice 
\cite{IB},
the \emph{J1-J2-J3} antiferromagnetic Heisenberg model \cite{j1-j2-j3}, 
and certain tetramer spin cluster Hamiltonians \cite{pal}. In general, the RVB spin liquid state can be expressed in terms of the Rokhsar-Kivelson Hamiltonians \cite{RK}.

\noindent\emph{Characterizing multisite entanglement.--}A multiparty pure state is said to be genuinely multiparty entangled if it is entangled across every possible bipartition of the system. 
%A quantitative measure of multisite entanglement is the geometric measure of entanglement which is the
% optimized distance between a genuine multisite 
%entangled state and the set of all states that are not genuinely multisite entangled. 
Examples of genuinely multisite entangled states include  the Greenberger-Horne-Zeilinger \cite{GHZ} and W \cite{W} states. %\textcolor{red}
{There also exist large superpositions of quantum spin-1/2 dimer states that are not genuinely multisite entangled \cite{j1-j2-j3, SS}.}
%For the isotropic square spin-1/2 lattice, with nearest neighbor dimer interaction, 
The following theorem shows that the state in Eq. (\ref{RVB}) is genuinely multisite entangled.
%We qualitatively study the multisite entanglement properties for the square spin-1/2 lattice state (\ref{RVB}), with dimer coverings and an isotropic interaction function $h_k(i,j)$. We show that such spin lattices are 
%genuinely multisite entangled.\\

\noindent \textbf{Theorem:} \emph{The pure state formed by superpositions of dimer coverings is
genuinely multisite entangled for all isotropic spin-1/2 lattices of arbitrary dimensions that are periodic or infinite in
all directions and all covering functions that are isotropic over the lattice.}

\noindent \texttt{Proof.} The superposition state of the spin-1/2 lattice consisting of \(2N\) particles, in Eq. (\ref{RVB}), is a pure state. 
To prove that this superposed state is genuinely multisite entangled, we are required to prove that the partial density
 matrix of the state across any bipartition cannot be pure. In other words, the density matrix of any $p$ spins, formed by tracing the remaining $2N-p$ spins, 
%of the bipartition 
is always mixed and, hence entangled 
to the rest. We conveniently divide the proof into the two cases where the number of spins (a) is finite in at least one part of the bipartition and (b) is infinite in both the parts.
%in the bipartition is a) finite and b) infinite.

\textit{a) Finite case:} For a rotationally invariant state, such as the dimer-covered spin-1/2 state under consideration, it is known that the partial density matrix of an arbitrary number of spins is also rotationally invariant. 
%Let us 
%denote this partial density matrix, corresponding to \(|\psi\rangle\), as \(\rho^{(X)}\).
Moreover, for an odd number of spin-1/2 particles, there is no pure quantum state that is rotationally invariant. Hence any odd bipartition of the system is always entangled to the rest of the system. 
For example, any single-site density matrix is
%the $X=1$ spin state has a density matrix 
%$\rho^{(1)}=
$\frac{1}{2} \mathcal{I}$, where $\mathcal{I}$ is the \(2\times 2\) identity matrix, and therefore is maximally entangled to the rest of the lattice.

Let us now consider the case of a bipartition with an even (finite) number of spins in one part. Consider any set \(X\) of an even number of sites. Let the partial density matrix of these sites in \(X\), corresponding to the 
state \(|\psi\rangle\), be \(\rho^{(X)}\).
Let us assume that \(\rho^{(X)}\) is pure, which would imply that \(|\psi\rangle\) is separable,  contrary to the statement of the theorem. 
Let \(X=X{'} \cup c\), where \(c\) contains an arbitrary but fixed odd number of sites \(< |X|\). (\(|S|\) denotes the cardinality of the set \(S\).) In particular, \(|c|\) can be unity. 
%For an even number of spins, suppose that let us consider $X+c$ spins in the partition that is pure, in contrast to the theorem. 
For an isotropic lattice, we can always find another equivalent set of spins, $Y{'} \cup c$, such that \(|Y{'}| = |X{'}|\), which is again pure (by the assumption).
% and has one spin ($c$) 
%that overlaps with $X+c$ pure state. 
%The density matrix of the $X$, $Y$ spins and $c$ spin can be written as $\rho^{(X)}$, $\rho^{(Y)}$ and $\rho^c$, respectively. 
%The corresponding von Nuemann entropies can be written 
%as $S(\rho^{(X)})$, $S(\rho^{(Y)})$ and $S(\rho^c)$. 
The strong subadditivity of von Neumann entropy \cite{strongadd} implies
%for any tripartite system between A, B, and C implies that $S(\rho^A)+S(\rho^B) \leq S(\rho^{AC})+S(\rho^{BC})$. 
%Hence we have, considering A=$X$, B=$Y$ and C=$c$,  
\begin{equation}
S(\rho^{(X{'})})+S(\rho^{(Y{'})}) \leq S(\rho^{(X{'} \cup c)})+S(\rho^{(Y{'} \cup c)}), 
\label{VN}
\end{equation}
where \(S(\cdot)\) denotes the von Neumann  entropy of its argument.
Since \(\rho^{(X)}\) and \(\rho^{(Y)}\) are pure, and since von Neumann entropy is non-negative, we have 
\(S(\rho^{(X{'})})+S(\rho^{(Y{'})})=0\), which in turn implies that \(\rho^{(X{'})}\) and \(\rho^{(Y{'})}\) are pure. This immediately implies that \(\rho^{(c)}\) is pure. 
This is a contradiction as \(|c|\) is odd. 
%
% Now, we have assumed $X+c$ and $Y+c$ is pure and hence the the right hand of (\ref{VN}) is 0. This implies that left hand of side of (\ref{VN}) is also 0, which is satisfied only when $X$ and $Y$ is itself pure. Hence for $X+c$ to be pure, 
%$c$ must be pure which we know is not true, since $\rho^{(c=1)}=1/2~ \mathcal{I}$. Hence, we cannot find any partition of the $2N$ spin-1/2 lattice that is pure and unentangled with the rest of the lattice sites.
This part of the proof was partially presented in Ref. \cite{asen_PRL}.

\textit{b) Infinite case:} 
%The proof can be further extended to the infinite case. 
Let us begin with the case of an infinite 2D square lattice partitioned into two half-planes by an infinite horizontal line. Let us assume that the reduced states are pure, contrary to the statement of the theorem. 
Let \(H_P\) denote the sites in one such half-plane. Consider now the set \(H_P \cup L_H\), where \(L_H\) is the infinite horizontal strip of sites with single-site width and that is directly adjacent to \(H_P\). 
By isotropy, if \(\rho^{(H_P)}\) is pure, \(\rho^{(H_P \cup L_H)}\) is also pure. Consequently, we can again use strong subadditivity to show that \(\rho^{(L_H)}\) is pure. 
Now if \(\rho^{(L_H)}\) is pure, by isotropy, \(\rho^{(L_V^{(p_1,p_2)})}\) is also pure, where \(L_V^{(p_1,p_2)}\) is the infinite vertical strip of sites with single-site width and that is obtained from \(L_H\) by rotating it by \(\pi/2\) radians
around the site with Cartesian coordinates \((p_1,p_2)\). Again using strong subadditivity, we find that \(\rho^{(p_1,p_2)}\) is pure, which is a contradiction. A similar proof works for boundaries that are not straight lines. 

One needs a separate
proof for the case when we want to prove that the infinite horizontal strip \(L_{H^r}\) having a width of \(r\) sites is not a product with the remaining portion of the lattice.  Assume, if possible, that \(\rho^{(L_{H^r})}\) is pure. Then,
by isotropy, \(\rho^{(L_{V^r}^{(p_1,p_2)})}\) is also pure, where \(L_{V^r}^{(p_1,p_2)}\) is obtained from \(L_{H^r}\) by a \(\pi/2\) rotation around \((p_1,p_2)\). Again applying strong subadditivity, we have that the reduced state of the 
\(2r\) spins in \(L_{H^r} \cap L_{V^r}^{(p_1,p_2)}\) is pure, which is a contradiction (by the (a) part of the proof). Similar proofs are possible for other infinite strips. 
Just like the proof of the (a) part,  the (b) part also works, with suitable modifications, for arbitrary isotropic lattices of arbitrary dimensions, e.g. the triangular and hexagonal lattices in 2D and the cubic lattice. 
% %
% 
% Suppose we divide the $2N=M.M'$ ($N \to \infty$) lattice along an infinite horizontal line dividing the lattice into two equivalent infinite pure 
% lattices ($N$). Using the inequality (\ref{VN}), we know that we can isolate an infinite horizontal strip of spins ($M\times 1$) that will be pure. In an isotropic lattice, a similar vertical strip of spins ($1\times M'$) can be obtained that is pure. 
% The two strips intersect at a site $c$. By enforcing (\ref{VN}), we can show that $c$ also must be pure, which again violates the rotational property of a single site state. Hence, no infinite bipartitions can be pure or truly unentangled in an isotropic lattice.
\hfill \(\blacksquare\)

%\section{Quantifying Multisite entanglement}
%\label{5:13}

\noindent\emph{Quantifying multisite entanglement.--}
%In the preceding section, 
%We have demonstrated that the dimer-covering superposition on an arbitrary isotropic lattice is genuinely multisite entangled. 
%It is interesting to quantify the amount of genuine multipartite entanglement present in such states. 
%This is important because 
The qualitative answer in the preceding section only indicates the nonvanishing quality of the multipartite entanglement content of the dimer-covering superposition on an arbitrary isotropic lattice. 
However, such an entanglement content can asymptotically graze to zero with increasing system size. Below, we show that this is not the case. Specifically, we provide an analytical method to quantify genuine multipartite entanglement 
of such dimer-covered superposition states and then show that the state under consideration possesses a relatively high genuine multiparty entanglement for arbitrarily large system size. 
To quantify the genuine multisite entanglement in an isotropic spin-1/2 lattice, with dimer coverings, we use 
a form of a geometric measure, that can be conveniently computed, called the generalized geometric measure (GGM) \cite{GGM} [for details, see the Supplemental Material \cite{supple}]. The GGM ($G(|\psi_R\rangle)$) is defined as
%\begin{equation}
$G(|\psi_R\rangle)=1- \max | \langle \chi|\psi_R\rangle|^2$
%\Lambda_{max}^2(|\psi_R\rangle)
%\end{equation}    
%where $\Lambda_{\max} (|\psi_R\rangle ) = \max | \langle \chi|\psi_R\rangle |$. 
where the maximization is performed over all possible $|\chi\rangle$ values that are not genuinely multisite entangled. 
%The maximization is over all possible $|\chi\rangle$ states.
%
%
%%%The GGM of an $R$-party quantum state is the optimized fidelity distance of the state $|\psi_R\rangle$ from the set of all states that are not genuinely multiparty entangled. 
%%More specifically, the GGM ($G(|\psi_R\rangle)$) can be calculated as
%%\begin{equation}
%%G(|\psi_R\rangle=1-\Lambda_{max}^2(|\psi_R\rangle)
%%\end{equation}    
%%where $\Lambda_{\max} (|\psi_R\rangle ) = \max | \langle \chi|\psi_R\rangle |$. $|\chi\rangle$ is an $R$-party quantum state with no genuine multisite entanglement. The maximization is over all possible $|\chi\rangle$ states.
%%
%%The GGM \cite{GGM} of an $R$-party pure state can be efficiently calculated by using the relation
%%\begin{equation}
%%G (|\psi_R \rangle ) =  1 - \max \{\lambda^2_{ K: L} |  K \cup  L = \{A_1,\ldots, A_R\},  K \cap  L = \emptyset\},
%%\end{equation}
%%% \end{equation}
%%%1 - \max\{e_{i:\textit{rest}}^{max} , e^{max}_{ij:\textit{rest}} \cdots |i \neq j\} , 
%%%\end{equation}
%%where \(\lambda_{K:L}\) is  the maximal Schmidt coefficients in all possible bipartite splits \(K: L\) of \(| \psi_R \rangle\). 
%The GGM can therefore be calculated by obtaining the reduced density matrices of the pure quantum state in all possible partitions. 
%Hence, the quantitative measure of multisite entanglement in nearest neighbor dimer spin-1/2 lattice can be characterized using the GGM.  
%
%%\subsection{Generating the reduced matrices}
%
%%The characterization of
The computation of the GGM
%, therefore, 
%Multisite entanglement is dependent 
depends 
on the efficient generation of arbitrary reduced density matrices across all possible bipartitions of the spin system. 
Certain properties such as two-point correlations, energy, and partition function, of such large superpositions, if found to be ground states of a Hamiltonian, can be evaluated in some cases by using approximate methods such as
mean field approximations and renormalization techniques \cite{DMRG}. 
%The limitations in computing all possible density matrices and its correlation properties for large-size quantum spin systems can be overcome by using 
We propose an iterative method, the \emph{density matrix recursion method} (DMRM) for 2D square spin-1/2 lattices, with NN dimer coverings [i.e. \(h_k(i_a,i_b) =1\) for NNs, and vanishing otherwise],
by which the limitations in computing all possible density matrices and the correlation properties of such lattice states can be overcome. 
%, that can generate reduced density matrix of large size spin systems using 
The method proceeds by deriving an algebraic recursion relation between the quantum states of small-sized spin systems that can be exactly computed. 
%The reduced density matrix generated can then be used to calculate the multisite entanglement as shown in the later sections. 
%
%We use the DMRM to calculate the reduced density matrices of the isotropic spin-1/2 spin lattice with nearest neighbor dimer interaction. 
We now apply the DMRM to two kinds of 2D systems: (a) the 
%\textcolor{red}
{\emph{perfect}} square lattices with an equal even number of sites on the horizontal and vertical sides and 
(b) the %\textcolor{red}
{\emph{imperfect}} square lattices having an even number of sites on the horizontal side but an odd number on the  vertical one. 
Note that the total number of spin sites is always even. 
%As we see later, 
%It is shown later that 
The difference between the behavior of GGM for perfect and imperfect square lattices is significantly reduced as the lattice size is increased.  
%
%\subsection{Perfect Square Lattice}
%\label{perf}

\noindent\emph{Perfect square lattice.--}We consider a spin-1/2 square lattice with $M$ spins along the horizontal and $M'$ spins in the vertical sides, such that $M \times M'$ (=$2N$) is the total number of spins. 
For a perfect lattice with even $M$ ($M'= M$), the spin state of the system, which we rename as $|M,M\rangle$, can be generated by using a recursion of two smaller sized imperfect spin states $|M-1,M\rangle$ and $|M-2,M\rangle$. 
We drop the second $M$ (=$M'$) from the states, as it remains unchanged. For a perfect lattice system with $M=M'=\mathcal{N}+2$, the recursion relation with open boundary condition can be written as
%\textcolor{blue}{
\begin{eqnarray}
\label{nonper}
|\mathcal{N}+2 \rangle &=& |\mathcal{N}+1\rangle |1\rangle_{n+2} + |\mathcal{N}\rangle |\bar{2}\rangle_{n+1,n+2}\nonumber\\
&=&|\mathcal{N}\rangle |2\rangle_{n+1,n+2} + |\mathcal{N}-1\rangle|\bar{2}\rangle_{n,n+1}|1\rangle_{n+2},
\end{eqnarray}
%}
where the subscripts correspond to the numbering of the sites on the horizontal side on the lattice. 
For our analysis, we consider a spin-1/2 square lattice state, with NN dimers, that is periodic along the horizontal axis \cite{Fan} (cf. Ref.\,\cite{DMRM}). 
%For nearest neighbor interaction, the isotropic function $h_k(i,j)$=1 for $(i,j)$ nearest 
%neighbor covering and zero for all other possible coverings. 
%Such a periodic lattice, enables us to recursively generate reduced density matrices of large spin systems using density matrix recursion method (DMRM) \cite{DMRM}. 
%The periodicity 
%accounts for all possible dimer coverings in a square spin-1/2 lattice.
Here, $|1\rangle_i$ refers to the state corresponding to ($M, M'$) = (1, $\mathcal{N}+2$) at the column $i$, and $|2\rangle_{i,j}$ refers to the state with
($M, M'$) = (2, $\mathcal{N}+2$) at the columns $i$ and $j$. The state $|\bar{2}\rangle_{i,j}$ is $|2\rangle_{i,j}-|1\rangle_i|1\rangle_j$. The subtraction in the term $|\bar{2}\rangle_{i,j}$ removes a repetition in the recursion. 
The recursion can be extended to states with the periodic boundary condition as 
$
\label{periodic}
|\mathcal{N}+2 \rangle_P = |\mathcal{N}+2\rangle_{1,n+2} + |\mathcal{N}\rangle_{2,n+1}|\bar{2}\rangle_{n+2,1}.
$
The states without subscript $P$ imply nonperiodic states. 
The recursion for the corresponding density matrix is given by
%The recursion can then be extended to the finite size denity matrix of the isotropic $M=\mathcal{N}+2$ spin system.
\begin{eqnarray}
\label{rho}
\rho^{(\mathcal{N}+2)}_P &=& \rho^{(\mathcal{N}+2)}+|\mathcal{N}\rangle\langle \mathcal{N}|_{(2,n+1)}|\bar{2}\rangle \langle \bar{2}|_{(1,n+2)}\nonumber\\
%|\mathcal{N}+2\rangle \langle \mathcal{N}+2|
&+&(|\mathcal{N}+2 \rangle \langle \mathcal{N}|_{2,n+1} \langle \bar{2}|_{n+2,1} + H.c.)
%&+&|\mathcal{N}\rangle\langle \mathcal{N}|_{(2,n+1)}|\bar{2}\rangle \langle \bar{2}|_{(1,n+2)}.
\end{eqnarray}
%}
where $ \rho^{(\mathcal{N}+2)}=|\mathcal{N}+2\rangle \langle \mathcal{N}+2|$ is the density matrix for the non-periodic recursion. Hence, using the relation for $\rho^{(\mathcal{N}+2)}_P$, we can generate the reduced density matrices across different partitions that can be used to 
%calculate the global entanglement properties of the spin system. Tracing over the entire spin system, barring the spins 
find the behavior of GGM in the system. Tracing over all the spins except the spins 
at columns $n+1$ and $n+2$, 
%so as to 
we obtain the corresponding  reduced density matrix
% of system size (2, $\mathcal{N}$+2) we obtain, for the non periodic case,
in the case of open boundary condition as
%\begin{eqnarray}
%\label{np}
$\rho^{(2)}_{(n+1,n+2)}= \mathcal{Z}_{\mathcal{N}} |2\rangle \langle 2|_{(n+1,n+2)} + 
\mathcal{Z}_{\mathcal{N}-1} \bar{\rho}_{n+1} \otimes |1\rangle\langle 1|_{(n+2)} 
+ (|2 \rangle_{n+1,n+2} \langle 1|_{n+2} \langle \xi_{\mathcal{N}} |_{n+1} + h.c.)$,
%\end{eqnarray}
where $\mathcal{Z}_{\mathcal{N}}= \langle\mathcal{N}|\mathcal{N}\rangle$, 
$\bar{\rho}_{n+1} = \Tr_n \large[|\bar{2}\rangle \langle \bar{2}|_{(n,n+1)}\large]$,
 and 
$
\langle\xi_{\mathcal{N}}|_{n+1}= \langle \bar{2}|_{n,n+1}\langle \mathcal{N}-1|\mathcal{N}\rangle.
$
%Extending the trace to the periodic ladder state, we get
%\textcolor{blue}
The reduced state, for periodic boundary condition, at columns $(n+1,n+2)$ is 
\begin{eqnarray}
\rho^{(2)}_{P}&=&\rho^{(2)}_{n+1,n+2} + \Tr_{1..n}[|\mathcal{N}\rangle \langle\mathcal{N}|_{(2,n+1)}|\bar{2}\rangle \langle\bar{2}|_{(1,n+2)} \nonumber\\
&+& (|\mathcal{N}\rangle_{2,n+1}|\bar{2}\rangle_{1, n+2}\langle \mathcal{N}+2| + \mathrm{H.c.})]\nonumber\\
&=& \rho^{(2)}_{n+1,n+2} + \beta^{(2)}_{1(n+1,n+2)} + (\beta^{(2)}_{2(n+1,n+2)} + \mathrm{H.c.}), \mathrm{where}\nonumber\\
\beta^{(2)}_{1} &=& \mathcal{Z}_{\mathcal{N}-1} |1\rangle \langle 1|_{(n+1)} \otimes  \bar{\rho}_{n+2} +\mathcal{Z}_{\mathcal{N}-2} \bar{\rho}_{n+1} \nonumber\\ %\mathcal{Z}_{\mathcal{N}-2} \bar{\rho}_{n+1} \otimes \bar{\rho}_{n+2}
&\otimes& ~\bar{\rho}_{n+2}+ (|1\rangle \langle \xi_{\mathcal{N}-1}|_{(n+1)} \otimes \bar{\rho}_{n+2} + h. c.),\nonumber  \mathrm{and}\\
%&+& (|1\rangle \langle \xi_{\mathcal{N}-1}|_{(n+1)} \otimes \bar{\rho}_{n+2} + h. c.),
\beta^{(2)}_{2} &=& |2\rangle_{n+1,n+2}\langle 1|_{n+1}\langle \xi_\mathcal{N}|_{n+2} + |2\rangle_{n+1,n+2} \nonumber\\
&\times& \textstyle{\sum_{i=1}^{\mathcal{N}}}\langle \mathcal{F}_i|_{n+2}\langle \xi_{\mathcal{N}-i}|_{n+1}+\bar{\rho}_{n+1}\otimes|1\rangle_{n+2}\langle \xi_\mathcal{N}|_{n+2}\nonumber\\
&+& \textstyle{\frac{1}{\mathcal{Z}_1}}(|\mathcal{F}_1\rangle_{n+1}|1\rangle_{n+2}\langle 1|_{n+1}\textstyle{\sum_{i=1}^\mathcal{N}}\langle \mathcal{F}_i|_{n+2}\mathcal{Y}^1_{\mathcal{N}-1}),
\end{eqnarray}
where $\mathcal{Y}^1_{\mathcal{N}}=\langle \mathcal{N}|\mathcal{N}-1\rangle|1\rangle$. 
%The states $|\mathcal{F}_i\rangle$ can be recursively generated, as shown below. To obtain the reduced density matrix for a large spin system, the parameters that appear upon tracing must be recursively generated from the parameters corresponding to smaller systems that can be exactly calculated. 
%
The different inner products can be calculated as follows. The recursion begins with evaluating
$\langle 1|\bar{2}\rangle =\sum_{i=1}^k \alpha^1_i |\alpha_i \rangle$, with $\{|\alpha_i\rangle\}_i$ forming an independent set of vectors consisting of 
% basis from all possible 
certain 
singlet combinations of an ($1,\mathcal{N}+2$)  spin system, where \(k\) is numerically caluclated; e.g., $|\alpha_1\rangle$=$|1\rangle$.  In general, we can write 
%\begin{equation}
%\label{inn}
$_n\langle \alpha_j|\bar{2}\rangle_{n,n+1} =(-1)^{n-1}\sum_i \alpha^j_i |\alpha_i \rangle_{n+1}$.
%\end{equation}
%\textcolor{blue}
Using this relation, we generate the inner product recursions:
\begin{eqnarray}
\mathcal{Z}_\mathcal{N}&=& \mathcal{Z}_1\mathcal{Z}_{\mathcal{N}-1}+\mathcal{Z}'_2\mathcal{Z}_{\mathcal{N}-2}+ 2 (-1)^{n-1}\textstyle{\sum_i} \alpha^1_i \mathcal{Y}^i_{\mathcal{N}-1},\nonumber \\
\mathcal{Y}_\mathcal{N}^j &=& _n\langle \alpha_j|(_{1,n-1}\langle\mathcal{N}-1|\mathcal{N}\rangle_{1,n})\nonumber\\
&=& \mathcal{A}_{j1}\mathcal{Z}_{\mathcal{N}-1}+ (-1)^{n-1} \textstyle{\sum_i} \alpha^j_i \mathcal{Y}^i_{\mathcal{N}-1}, ~~\mathrm{and}\nonumber\\
\langle \xi_\mathcal{N}|_{n+1}&=&\langle \bar{2}|_{n,n+1}\langle \mathcal{N}-1|\mathcal{N}\rangle = \textstyle{\sum_{i=1}^{\mathcal{N}}} \mathcal{Z}_{\mathcal{N}-i}\langle \mathcal{F}_i|_{n+1},
\end{eqnarray}%\textcolor{red}
%%where $\mathcal{Z}'_2=\langle \bar{2}|\bar{2}\rangle$ and $\mathcal{A}_{ij}= \langle\alpha_i|\alpha_j\rangle$. Similarly, the other states can be generated using subsequent recursions,
{where $\mathcal{Z}'_2$ and $\mathcal{A}_{ij}$ are $\langle \bar{2}|\bar{2}\rangle$ and $\langle\alpha_i|\alpha_j\rangle$, respectively.}
%\begin{eqnarray}
%\langle \xi_\mathcal{N}|_{n+1}&=&\langle \bar{2}|_{n,n+1}\langle \mathcal{N}-1|\mathcal{N}\rangle = \sum_{i=1}^{\mathcal{N}} \mathcal{Z}_{\mathcal{N}-i}\langle \mathcal{F}_i|_{n+1},
%\end{eqnarray}
%&=&\sum_{i=1}^{\mathcal{N}} \mathcal{Z}_{\mathcal{N}-i}\langle \mathcal{F}_i|_{n+1}
$\langle \mathcal{F}_{i}|_{n+1}=~_{n,n+1}\langle \bar{2} |\mathcal{F}_{i-1}\rangle_n$ and $|\mathcal{F}_0\rangle_{n}=|1\rangle_{n}$. On expanding, we can write
$
\langle \mathcal{F}_{i}|_{n+1}=~_{n,n+1}\langle \bar{2} |\mathcal{F}_{i-1}\rangle_n= \sum_k g_k^i\langle \alpha_k|_{n+1},
$
where $g_j^i = \sum_k g_k^{i-1} \alpha^k_j$. Hence all the terms can be recursively calculated provided the parameters for small spin systems can be accurately estimated. 
The terms $\alpha_i^j$($i,j=$ 1 to $k$), of $\mathcal{Z}_1$, $\mathcal{Z}'_2$, and $\mathcal{A}_{ij}$ need to be exactly calculated. The value of $k$ needs to be determined for a small system size by solving the linear equation system in $_n\langle \alpha_j|\bar{2}\rangle_{n,n+1}$.

%\subsection{Imperfect Square Lattice}
%\label{imperf}

\noindent\emph{Imperfect square lattice.--} Let us now consider spin lattices with \(M\) sites on the horizontal side and \(M{'}\) on the vertical one, where \(M\) is even, while \(M{'}\) is odd. Here  $M \times M'$ (=$2N$) and $M'$ (=$M \pm 1$).
% on the vertical side is odd. 
For a system containing $M=\mathcal{N}+2$ spin sites along the horizontal side, the recursion of the periodic state $|\mathcal{N}+2,\mathcal{N}+2 \pm 1\rangle_P$ can be written as 
%\begin{eqnarray}
%\label{operiodic}
$
|\mathcal{N}+2\rangle_P = |\mathcal{N}\rangle_{1,n} |2\rangle_{n+1,n+2} + |\mathcal{N}\rangle_{2,n+1} |2\rangle_{n+2,1}.\,\,\,
$
%\end{eqnarray}
Here, the constant $M'=\mathcal{N}+2 \pm 1$ has been omitted from the notation of the state. 
$|2\rangle_{i,j}$ refers to the state corresponding to ($M, M'$) = (2, $\mathcal{N}+2 \pm 1$). 
%The recursion relation is much simpler compared to the case of perfect isotropic lattices due to the absence of $|1\rangle$ in the odd ladder.

%The recursion of the density matrix, obtained using (\ref{operiodic}), is given by
%$\rho^{N+2}_P = |\mathcal{N}+2\rangle \langle \mathcal{N}+2|_P$. 
%Similarly, 
The reduced density matrices can be recursed by using the recursion relation of $|\mathcal{N}+2\rangle_P$.
%Eq. (\ref{operiodic}). 
In particular, we  trace out the spins in the columns ranging from 1 to \(n\), for obtaining a reduced density matrix of the spins at columns ($n+1, n+2$):
%, to obtain a state of size $(2,\mathcal{N}+1)$.
%\textcolor{blue}{
\begin{eqnarray}
\rho^{(2)}_{P}&=& \Tr_{1...n}[|\mathcal{N}+2\rangle \langle \mathcal{N}+2|_P]\nonumber\\
%&=& \Tr_{1..n}[|\mathcal{N}\rangle \langle \mathcal{N}|_{1,n}|2 \rangle \langle 2|_{n+1,n+2} 
%+ |\mathcal{N}\rangle \langle \mathcal{N}|_{2,n+1}|2 \rangle \langle 2|_{n+2,1} \nonumber\\
%&+& (|\mathcal{N}\rangle_{1,n}|2\rangle_{n+1,n+2}\langle \mathcal{N}|_{2,n+1}\langle 2|_{1,n+2} + h. c.)]\nonumber\\
&=&\mathcal{Z}_\mathcal{N} |2\rangle \langle 2|_{(n+1,n+2)} + \mathcal{Z}_{\mathcal{N}-2} \bar{\rho}_{n+1} \otimes \bar{\rho}_{n+2}\nonumber\\
&+& (|2\rangle_{n+1,n+2}\langle \chi_\mathcal{N}|_{n+1,n+2} + \mathrm{H.c.}),
\end{eqnarray}
%After simplification, the above equation reads
%This can be rewritten as
%e expression for the traced reduced density matrix:
%\begin{eqnarray}
%\label{p}
%$\rho^{(2)}_{P(n+1,n+2)} =\mathcal{Z}_\mathcal{N} |2\rangle \langle 2|_{(n+1,n+2)} + \mathcal{Z}_{\mathcal{N}-2} \bar{\rho}_{n+1} \otimes \bar{\rho}_{n+2} + (|2\rangle_{n+1,n+2}\langle \chi_\mathcal{N}|_{n+1,n+2} + h. c.)$,
%\end{eqnarray}
where 
%the normalization is defined as 
\(\mathcal{Z_N} 
%= \langle \mathcal{N}| \mathcal{N}\rangle 
= \mathcal{Z}_2^{N/2}\).
The recursion for the state $|\chi_\mathcal{N}\rangle$ is given by
%\begin{eqnarray}
$\langle \chi_\mathcal{N}|_{n+1,n+2}=\langle 2|_{1,n+2}\langle \mathcal{N}|_{2,n+1}|\mathcal{N}\rangle_{1,n}
= \langle 2|_{1,n+2} \langle 2|_{2,3}...\langle 2|_{n,n+1}|2 \rangle_{1,2}|2 \rangle_{3,4}...|2\rangle_{n-1,n}$.
%\end{eqnarray}
The last recursion can be done numerically to solve for $\langle 2|_{i,i+3}\langle 2|_{i+1,i+2}|2 \rangle_{i,i+1}$.
%
%We observe that the recursion relation of the imperfect isotropic lattices is simpler than the recursion for perfect lattices. 
%We are interested in observing the properties of multiparty entanglement of the square spin-1/2 lattice, 
%
In the succeeding subsection, finite-size calculations shows that 
the behavior of multisite entanglement of the two types of lattices merges as the size of the system 
increases.
% which can be confirmed in the succeeding section. 
%For large square lattice of dimensions ($\mathcal{N},\mathcal{N}$), 
%the global 
%it is natural to assume that genuine multisite entanglement will be close to that for a lattice of dimensions ($\mathcal{N},\mathcal{N} \pm 1$). 

We have checked for small system size, that the square lattice under consideration is long-range entangled, in the sense introduced in Ref. \cite{long-range}. Hence, genuine multisite entanglement can be used as an indicator of topological long-range order, for the RVB liquid states. Moreover, noisy admixtures of these superposed dimer states can be shown to be differential local convertible, implying that such states are deep inside a macroscopic phase of the corresponding quantum many-body system \cite{DLC}.
%\subsection{Finite size scaling}
%\label{fss}

%The multisite entanglement of the nearest neighbor dimer spin-1/2 lattice can be calculated using GGM on the reduced density matrices obtained using DMRM. We use the characterization to generate results for some perfect and imperfect square spin-1/2 lattices. We observe that for increasing system size, the global entanglement of the system, characterized by its multisite entanglement (GGM), merges for the perfect and the imperfect square lattices. 
\noindent\emph{Finite-size scaling.--}Let us now calculate the genuine multipartite entanglement of spin-1/2 square lattices
by using the DMRM for such lattices. We find the GGM for perfect as well as imperfect square  lattices and observe that for increasing system size, 
the GGM converges to the value 0.358 (Fig.\,1).
%saturates for the perfect and the imperfect square lattices.
%
\begin{figure}
\begin{center}
\epsfig{figure = 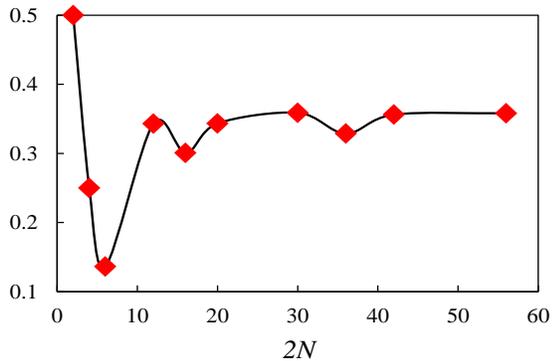, height=.2\textheight, width=0.4\textwidth,angle=-0}
\caption{(Color online.) The behavior of GGM in the case of a square spin-1/2 lattice, with increasing total number of spins ($2N$). 
%The total number of spins with positive integer root correspond to perfect square lattices. 
%The points corresponding to non-integer roots correspond to imperfect square lattices. We observe that the genuine multisite entanglement merges at large sized systems. The infinite 2D behavior of the square lattice can be obtained by finite size scaling. 
}
\label{fig:2}
\end{center}
\end{figure}
%
%%In Fig. 2, we plot the
%the genuine multisite entanglement, as measure by 
%%GGM with increasing total number of spins ($2N$). 
%The GGM converges as the system size is increased and the global entanglement property is indifferent to the perfect (points corresponding to $2N$ that has a positive integer root) or imperfect nature of the square lattice. 
Using finite-size scaling, the behavior of GGM for finite-sized lattices
can be used to estimate the GGM for an infinite square lattice.
%, having infinite sites. 
The scaling 
%estimates gives us an approximate expression for GGM with system size;
analysis gives 
\begin{equation}
G (|\psi\rangle )\approx G_c(|\psi \rangle )\pm k ~\mathbb{N}^{-x},
\label{scale}
\end{equation}
where $\mathbb{N}=2N$ and $G_c(|\psi \rangle )$ is an estimated value of GGM for the infinite lattice, based on the average of the last two values of GGM given in Fig.\,1, with  $k$ being a 
%scaling 
constant. 
The value of $x$, as estimated by finite-size scaling, using $G_c(|\phi_N \rangle )$=0.358, is $x=$1.82, whereas $k$=1.77.

%Hence, we observe that the large size GGM quickly to a finite value as shown by relation (\ref{scale}). We are thus, able to quantify an estimated value of the global entanglement of an infinite sized two-dimensional square lattice. 
%
%
%
%%\section{Discussions}
%\label{disc}

%We have shown in our work that the multisite entanglement in square isotropic spin-1/2 systems can be suitably characterized using a computable measure of genuine geometric multipartite entanglement. The large system properties can be derived from reduced density matrices of the system that can be effectively calculated using an iterative method to recursively generate these reduced matrices. The calculations enable us to exactly quantify and calculate the multisite entanglement in systems with relatively large number of field. These calculations can then be scaled using finite size scaling to estimate with significant precision the global entanglement of a two-dimensional infinite isotropic spin lattice. 

\noindent\emph{Discussion.--}We investigate the multipartite entanglement of a large superposed state consisting of dimer coverings of a spin-1/2 isotropic lattice. 
We first showed that the state, if isotropic over the lattice, with either periodic boundary conditions or is an infinite lattice, is genuinely multipartite entangled, regardless of the geometry and dimension of the isotropic lattice. 
To measure its multisite entanglement content, we have presented  a technique, which we have referred to as the density matrix recursion method for square lattices, 
 to analytically study arbitrary state parameters of an arbitrary number of sites, including its genuine multisite entanglement.
% of a largely superposed state in a square isotropic spin-1/2 lattice. 
The method was then employed to investigate the finite-size scaling behavior of the generalized geometric measure, which is a %\textcolor{red}
{computable measure of genuine multipartite entanglement
for finite-sized square lattices} that enables us to to estimate the genuine multisite entanglement of an infinite square lattice. %\textcolor{red}{We note that GGM is a computable measure of genuine multipartite entanglement.} 
%
%\textcolor{red}{An interesting property of the square lattices we consider are that they are long-range entangled \cite{long-range}. Hence, genuine multisite entanglement can be an indicator of topological long-range order, for the states we consider. Further, noisy admixtures of these large superposed dimer states are differential local convertible, implying that such states are deep inside a macroscopic phase of the corresponding quantum many-body system \cite{DLC}.}
%
Genuine multisite entanglement is potentially a basic ingredient in building large-scale quantum computers and also in implementation of multiparty quantum communication. 
The method presented can be a useful tool if such highly superposed systems are considered for performing quantum tasks. 
Specifically, the iterative method
can be employed 
to derive reduced density matrices that will in turn be fruitful in the calculation of nearest-neighbor bipartite entanglement as well as that of other two-point correlation functions.
% in many body theory. 
%This could prove useful in studying strongly correlated systems and quantum phase transitions. The characterization of multisite entanglement can be extended to other quantum systems for use in various quantum protocols.

%\begin{acknowledgments}
The work of H.S.D. is supported by the University Grants Commission. H.S.D. thanks the Harish-Chandra Research Institute (HRI) for hospitality and support during visits. 
We acknowledge computations performed at the cluster computing facility at HRI and the UGC-DSA facility at the Jawaharlal Nehru University.

%\end{acknowledgments}
  
\section*{Supplemental Material}

\subsection{The bipartite lattice with nearest-neighbor dimer coverings}

\noindent\emph{The spin state}: An isotropic spin-1/2 system, with dimer coverings, is conveniently defined by  using a \emph{bipartite lattice}. 
A bipartite lattice is one which is divided into two sublattices, 
A and B, such that each site
of sublattice A has only sites of sublattice B as its nearest neighbors (see Fig.\,2). Moreover, each lattice site is occupied by a spin-1/2 particle.

\begin{figure}[htb]
\label{fig:1}
%\begin{center}
\epsfig{figure = 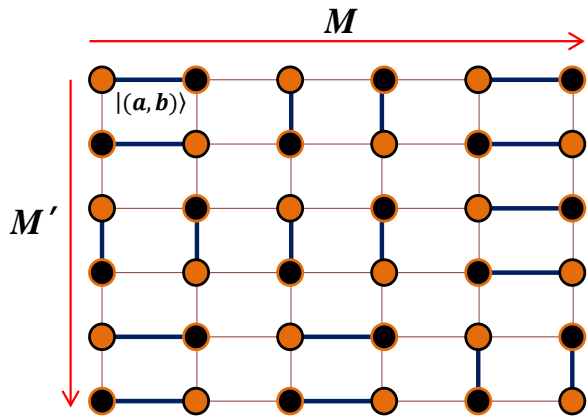, height=.23\textheight, width=0.43\textwidth,angle=-0}
\caption{(Color online.) A square spin-1/2 bipartite lattice with nearest-neighbor dimer coverings. Here, $M$, $M'$=6. 
%in the form of a bipartite lattice. 
The sublattice $A$ consists of the sites denoted by the lighter circles with darker borders,
% (yellow with black) 
while those in  $B$ has darker circles with lighter borders.
% (black with yellow). 
The thick solid lines  show the nearest-neighbor dimer states ($|(a_i,b_j)\rangle$) from a site in sublattice $A$ to another in $B$. The 
figure represents a possible dimer covering. The final state is the superposition of all such dimer coverings.}
%\end{center}
\end{figure}

% contains a spin-1/2 qubit and each qubit in sublattice A (B) can form a dimer with another qubit in sublattice B(A). 
The dimer-covering state on such a bipartite lattice can be 
%The spin-1/2 state of $2N$ spin system can then be 
written in the form \cite{anders},
\begin{equation}
|\psi\rangle=\sum_k h_k(i_a,i_b)|(a_1,b_1),(a_2,b_2)...(a_N,b_N)\rangle_k,
\label{RVB}
\end{equation}
where each $k$ refers to a single dimer covering ($\{|a_i,b_i\rangle\}_i$) on the entire lattice
%between sites of the two sublattices, 
with $N$ spins in each sublattice. 
Here, the ``covering function'' $h_k(i_a,i_b)$ is isotropic over the lattice sites $i_a \in A, i_b \in B$. 
%The final state is the superposition of all such possible coverings of the lattice and is rotationally invariant.

% For our analysis, we consider a spin-1/2 lattice, with nearest neighbor dimer interaction, that is periodic along the horizontal axis \cite{Fan}. For nearest neighbor interaction, the isotropic function $h_k(i,j)$=1 for $(i,j)$ nearest 
% neighbor covering and zero for all other possible coverings. Such a periodic lattice, enables us to recursively generate reduced density matrices of large spin systems using density matrix recursion method (DMRM) \cite{DMRM}. The periodicity 
% accounts for all possible dimer coverings in a square spin-1/2 lattice.

% and closely approximates an isotropic RVB state. 
%The DMRM reduces computation of to smaller system size that can be scaled along the non-periodic vertical axis thus giving us isotropic square spin-1/2 lattice states. The GGM allows us to quantify genuine multisite entanglement in these lattices. 

The above states, 
% formed by the superposition of 
with
%nearest neighbor 
short-range dimer coverings are often referred  as resonating valence bond (RVB) states \cite{super, bhaskar}. 
In quantum spin ladders, these states have been argued to be possible ground state approximations 
of certain Heisenberg spin-1/2 systems. 
%Nearest neighbor dimer interactions, similar to RVB, can be simulated using optical lattices \cite{opt} and photons \cite{photon} to generate quantum systems that harness the quantum correlation properties of such 
%states for use in information protocol. 
Spin systems, such as RVB ladders, have been extensively used to study entanglement properties 
\cite{EP}.

%\section{Characterizing multisite entanglement}
%\label{ME}

\subsection{Generalized geometric measure}

To quantify the genuine multisite entanglement in an isotropic spin-1/2 lattice, with dimer coverings, we use 
%a generalized form of a geometric measure, that can be conveniently computed, called 
the generalized geometric measure (GGM) \cite{GGM}.
The GGM of an $R$-party quantum state is the optimized fidelity distance of the state $|\psi_R\rangle$ from the set of all states that are not genuinely multiparty entangled. 
More specifically, the GGM ($G(|\psi_R\rangle)$) can be calculated as
\begin{equation}
G(|\psi_R\rangle=1-\Lambda_{max}^2(|\psi_R\rangle)
\end{equation}    
where $\Lambda_{\max} (|\psi_R\rangle ) = \max | \langle \chi|\psi_R\rangle |$. $|\chi\rangle$ is an $R$-party quantum state with no genuine multisite entanglement. The maximization is over all possible $|\chi\rangle$ states.

The GGM \cite{GGM} of an $R$-party pure state can be efficiently calculated by using the relation
\begin{equation}
G (|\psi_R \rangle ) =  1 - \max \{\lambda^2_{ K: L} |  K \cup  L = \{A_1,\ldots, A_R\},  K \cap  L = \emptyset\},
\end{equation}
% \end{equation}
%1 - \max\{e_{i:\textit{rest}}^{max} , e^{max}_{ij:\textit{rest}} \cdots |i \neq j\} , 
%\end{equation}
where \(\lambda_{K:L}\) is  the maximal Schmidt coefficients in all possible bipartite splits \(K: L\) of \(| \psi_R \rangle\). 
%The GGM can therefore be calculated by obtaining the reduced density matrices of the pure quantum state in all possible partitions. 
%Hence, the quantitative measure of multisite entanglement in nearest neighbor dimer spin-1/2 lattice can be characterized using the GGM.  

The GGM has been used as a measure of genuine multipartite entanglement in studying different aspects of quantum information theory \cite{GGM, GGM1} and in many body systems \cite{DMRM, GGM2}.


\begin{thebibliography}{99}
%\bibitem{etc-multi} etc-multi \#\#\#

\bibitem{horo} R. Horodecki \emph{et al.}, 
%P. Horodecki, M. Horodecki, and K. Horodecki, 
Rev. Mod. Phys. \textbf{81}, 865 (2009).



\bibitem{computation} H.J. Briegel \emph{et al.},
% D. E. Browne, W. D\"{u}r, R. Raussendorf and M. Van den Nest, 
Nat. Phys. \textbf{5}, 19 (2009).

 


\bibitem{communicate} 
%J. Kempe, Phys. Rev. A \textbf{60}, 910 (1999); V. Scarani and N. Gisin, Phys. Rev. Lett. \textbf{87}, 117901 (2001); 
%N. J. Cerf, S. Massar, and S. Schneider, Phys. Rev. A \textbf{66}, 042309 (2002); C. Kruszynska, S. Anders, W. D\"{u}r, and H. J. Briegel, Phys. Rev. A \textbf{73}, 062328 (2006); 
A. Sen(De) and U. Sen, Phys. News \textbf{40}, 17 (2010), arXiv:1105.2412.

\bibitem{crypto} M. \.Zukowski \emph{et al.}, 
%M. Horne, and H. Weinfurter,
Acta Phys. Pol. \textbf{93}, 187 (1998); 
M. Hillery \emph{et al.}, 
%V. Bu{\v z}ek, and A. Berthiaume, 
Phys. Rev. A \textbf{59}, 1829 (1999);
R. Demkowicz-Dobrzanski \emph{et al.},
% A. Sen(De), U. Sen, and M. Lewenstein, 
\emph{ibid.} \textbf{80}, 012311 (2009); 
N. Gisin \emph{et al.},
% G. Ribordy, W. Tittel, and H. Zbinden, 
Rev. Mod. Phy. \textbf{74}, 145 (2002).


\bibitem{qss} R. Cleve \emph{et al.},
% D. Gottesman, and H.-K. Lo, 
Phys. Rev. Lett. \textbf{83}, 648 (1999); 
A. Karlsson \emph{et al.},
% M. Koashi, and N. Imoto,
Phys. Rev. A \textbf{59}, 162 (1999).


\bibitem{QPT} 
T.-C. Wei \emph{et al.},
% D. Das, S. Mukhopadyay, S. Vishveshwara, and P. M. Goldbart, 
\emph{ibid.} \textbf{71}, 060305(R) (2005); T.R. de Oliveira \emph{et al.}, Phys. Rev. A \textbf{73}, 010305(R) (2006);
%G. Rigolin, M. C. de Oliveira,
%and E. Miranda, 
Phys. Rev. Lett. \textbf{97}, 170401 (2006);
%T.R. de Oliveira \emph{et al.}, 
%G. Rigolin and M. C. de Oliveira, 
%
G. Costantini \emph{et al.},
% P. Facchi, G. Florio, and S. Pascazio, 
J. Phys. A: Math. Theor. \textbf{40}, 8009 (2007);
%\textcolor{red}
{R. Or{\'u}s, Phys. Rev. Lett. \textbf{100}, 130502 (2008); R. Or{\'u}s \emph{et al.} \emph{ibid.} \textbf{101}, 025701 (2008); R. Or{\'u}s and T.-C. Wei, Phys. Rev. B \textbf{82}, 155120 (2010)};
D. Buhr \emph{et al.},
% M. E. Carrington, T. Fugleberg, R. Kobes, G.
%Kunstatter, D. McGillis, C. Pugh, D. Ryckman, 
J. Phys. A: Math. Theor. \textbf{44}, 365305 (2011); 
%M.N. Bera \emph{et al.},
% R. Prabhu, A. Sen(De), and U. Sen, 
%arXiv:1209.1523; A. Biswas \emph{et al.},
% R. Prabhu, A Sen(De), and U. Sen, 
%arXiv:1211.3241; %\textcolor{red}
{R. Or{\'u}s \emph{et al.}, arXiv:1304.1339}.




%T. R. de Oliveira, G. Rigolin and M. C. de Oliveira, Phys. Rev. A \textbf{73}, 010305(R) (2006); M. N. Bera, R. Prabhu, A. Sen(De), and U. Sen, arXiv:1209.1523 [quant-ph] (2012).

\bibitem{bio2} M. Sarovar \emph{et al.},
% A. Ishizaki, G.R. Fleming, and K.B. Whaley, 
Nat. Phys. \textbf{6}, 462 (2010); J. Zhu \emph{et al.},
% S. Kais, A. Aspuru-Guzik, S. Rodriques, B. Brock, and P. J. Love, 
J. Chem. Phys. \textbf{137}, 074112 (2012).



% \bibitem{loss} M.M. Wolf 
% %\emph{et al.},
% , F. Verstraete, M.B. Hastings and J.I. Cirac,
%  Phys. Rev. Lett. \textbf{100}, 070502 (2008).
 
%\bibitem{separable} R. Horodecki,  Phys. Lett. A \textbf{187}, 145 (1994).

%\bibitem{MS3} P. Facchi, G. Florio, and S. Pascazio, Phys. Rev. A \textbf{74}, 042331 (2006).


\bibitem{MS1} V. Coffman \emph{et al.},
 %J. Kundu, and W. K. Wootters, 
 Phys. Rev. A \textbf{61}, 052306 (2000); G. Vidal \emph{et al.},
 %W. D\"{u}r, and J. I. Cirac, 
 Phys. Rev. Lett. \textbf{85}, 658 (2000); 
J. Eisert and H.J. Briegel, Phys. Rev. A \textbf{64}, 022306 (2001); M. Horodecki \emph{et al.},
% P. Horodecki and R. Horodecki, 
Phys. Lett. A 283, \textbf{1} (2001); H. Barnum and N. Linden, J. Phys. A: Math. Theor. \textbf{34}, 6787 (2001); 
D.A. Meyer and N.R. Wallach, J. Math. Phys. \textbf{43}, 4273 (2002); 
D. Collins \emph{et al.},
 %N. Gisin, S. Popescu, D. Roberts and V. Scarani, 
 Phys. Rev. Lett. \textbf{88}, 170405 (2002); 
T.-C. Wei and P.M. Goldbart, Phys. Rev. A \textbf{68}, 042307
(2003);
A. Miyake \emph{ibid.} \textbf{67}, 012108 (2003); F. Verstraete \emph{et al.},
% J. Dehaene, and B. De Moor, 
\emph{ibid.} \textbf{68}, 012103 (2003); 
%F. Verstraete \emph{et al.},
% M. Popp, and J. I. Cirac, 
Phys. Rev. Lett. \textbf{92}, 027901 (2004); C.S. Yu and H.S. Song, Phys. Rev. A \textbf{72}, 022333 (2005); 
A. Osterloh and J. Siewert, \emph{ibid.} \textbf{72}, 012337 (2005);
P. Facchi \emph{et al.},
% G. Florio, and S. Pascazio, 
\emph{ibid.} \textbf{74}, 042331 (2006);
D.L. Deng \emph{et al.},
% Z. S. Zhou, and J. L. Chen, 
\emph{ibid.} \textbf{80}, 022109 (2009); 
P. Krammer \emph{et al.},
% H. Kampermann, D. Bru{\ss}, R. A. Bertlmann, L. C. Kwek and C. Macchiavello, 
Phys. Rev. Lett. \textbf{103}, 100502 (2009); 
M. Huber \emph{et al.}, 
%F. Mintert, A. Gabriel and B. C. Hiesmayr, 
\emph{ibid.} \textbf{104}, 210501 (2010);
Q.-Q. Shi \emph{et al.}, New J. Phys. \textbf{12}, 025008 (2010); %\textcolor{red}
{B.-Q. Hu {et al.} \emph{ibid.} \textbf{13} 093041 (2011)};
% Roman Orus, John Ove Fjaerestad, Huan-Qiang Zhou
J.-D. Bancal \emph{et al.},
% N. Brunner, N. Gisin, and Y.-C. Liang, 
Phys. Rev. Lett. \textbf{106}, 020405 (2011).




\bibitem{GGM} A. Sen(De) and U. Sen, Phys. Rev. A \textbf{81}, 012308 (2010);
%
%\bibitem{amadernext} A. Sen(De) and U. Sen, 
arXiv:1002.1253.

%\bibitem{organic} L. Pauling, Proc. R. Soc. London, Ser. A \textbf{196}, 343 (1949)


\bibitem{ander} %\textcolor{red}
{P.W. Anderson, Mater. Res. Bull. \textbf{8} 153 (1973). }

\bibitem{super} P.W. Anderson, Science \textbf{235}, 1196 (1987); E. Dagatto and T.M. Rice, \emph{ibid.} \textbf{271}, 618 (1996).



\bibitem{critical} M. Lewenstein \emph{et al.},
% \emph{et al.},
%A. Sanpera, V. Ahufinger, B. Damski, A. Sen(De), and U. Sen,
%Comments: Review article. v2: published version, 135 pages, 34 figures
Adv.  Phys. \textbf{56}, 243 (2007);
L. Amico 
\emph{et al.},
% R. Fazio, A. Osterloh, and V. Vedral, 
Rev. Mod. Phys. \textbf{80}, 517 (2008).

\bibitem{scal} E. Dagatto 
\emph{et al.},
%J. Riera, and D. Scalapino, 
Phys. Rev. B \textbf{45}, 5744 (1992);
%
S.R. White
\emph{et al.},
% R. M. Noack, and D. J. Scalapino, 
Phys. Rev. Lett. \textbf{73}, 886 (1994).



\bibitem{qc} A.Y. Kitaev, Ann. Phys. (Leipzig) \textbf{303}, 2 (2003).



\bibitem{opt} S. Nascimbne \emph{et al.}, Phys. Rev. Lett. \textbf{108}, 205301 (2012).

%%M. Greiner \emph{et al.}, 
%O. Mandel, T. Esslinger, T. W. Hansch and I. Bloch, 
%%Nature \textbf{415}, 39 (2002); F. Verstraete \emph{et al.},
% J. I. Cirac, and I. J. Latorre, 
%Phys. Rev. A \textbf{79}, 032316 (2009).

\bibitem{photon} X.-S. Ma \emph{et al.},
% B. Dakic, W. Naylor, A. Zeilinger, and P. Walther, 
Nat. Phys. \textbf{7}, 399 (2011).

\bibitem{book-scale} M.N. Barber, in \emph{Finite size scaling, in Phase Transitions
and critical phenomena}, ed. C. Domb and J. L. Leibovitz ( Academic Press, London, 1983).

\bibitem{anders} S. Liang
\emph{et al.},
% B. Doucot, and P. W. Anderson, 
Phys. Rev. Lett. \textbf{61}, 365 (1988).

\bibitem{supple} See Supplemental Material for a description of the bipartite spin lattice with nearest neighbor
dimer coverings and a discussion on the generalized geometric measure as a genuine multipartite entanglement
measure.

\bibitem{IB} %\textcolor{blue}
{I. Bose and A. Ghosh, Phys. Rev. B \textbf{56}, 3149 (1997).}

\bibitem{j1-j2-j3} %\textcolor{red}
{B. Kumar, Phys. Rev. B \textbf{66} 024406 (2002); M. Mambrini, \emph{et al.}, \emph{ibid.} \textbf{74}, 144422 (2006).}


\bibitem{pal} %\textcolor{red}
{A.K. Pal and I. Bose, J. Phys. B \textbf{44} 045101 (2011).}

\bibitem{RK} %\textcolor{red}
{D. Rokhsar and S.A. Kivelson, Phys. Rev. Lett. \textbf{61} 2376 (1988); J. Cano and P. Fendley, \emph{ibid.} \textbf{105}, 067205 (2010).}


%\bibitem{asen_PRL} A. Chandran, D. Kaszlikowski, A. Sen(De), U. Sen, and V. Vedral, Phys. Rev. Lett \textbf{99}, 170502 (2007).


\bibitem{GHZ} D.M. Greenberger \emph{et al.},
% M. A. Horne, and A. Zeilinger, 
in \emph{Bell’s Theorem, Quantum Theory, and Conceptions of
the Universe}, ed. M. Kafatos (Kluwer Academic, Dordrecht, 1989).


\bibitem{W} A. Zeilinger \emph{et al.},
% M. A. Horne, and D. M. Greenberger, 
in \emph{Proc. Squeezed States \& Quantum Uncertainty}, eds. D. Han, Y.S. Kim, and W.W. Zachary, NASA Conf. Publ. 3135 (1992);
%
%\bibitem{W1} 
W. D\"{u}r \emph{et al.},
% G. Vidal, and J. I. Cirac, 
Phys. Rev. A \textbf{62}, 062314 (2000).

\bibitem{SS} %\textcolor{red}
{B.S. Shastry and B. Sutherland, Physica (Amsterdam) \textbf{108B+C}, 1069 (1981).}

\bibitem{strongadd} E.H. Lieb and M.B. Ruskai, J. Math. Phys. \textbf{14}, 1938 (1973).
\bibitem{asen_PRL} A. Chandran \emph{et al.},
% D. Kaszlikowski, A. Sen(De), U. Sen, and V. Vedral, 
Phys. Rev. Lett \textbf{99}, 170502 (2007).
%\bibitem{mean} S. Gopalan, T. M. Rice, and M. Sigirst, Phys. Rev. B \textbf{49}, 8901 (1994).

\bibitem{DMRG} S.R. White, Phys. Rev. Lett. \textbf{69}, 2863 (1992); Phys. Rev. B \textbf{48}, 10345 (1993); S. Gopalan \emph{et al.},
% T. M. Rice, and M. Sigirst, 
\emph{ibid.} 
%Phys. Rev. B 
\textbf{49}, 8901 (1994); M. Roncaglia,
%\emph{et al.},
M. Greven
\emph{et al.},
%R.J. Birgeneau, and U.-J. Wiese, 
Phys. Rev. Lett. \textbf{77}, 1865 (1996); Y. Nishiyama \emph{et al.},
% N. Hatano and M. Suzuki, 
J. Phys. Soc. Jpn. \textbf{65}, 560 (1996);
G. Sierra and M.A. Martin-Delgado, 
Phys. Rev. B \textbf{56}, 8774 (1997); G. Sierra and M.A. Martin-Delgado, 
\emph{ibid.} \textbf{60}, 12134 (1999).

\bibitem{Fan} Y. Fan and M. Ma, Phys. Rev B \textbf{37}, 1820 (1988).

\bibitem{DMRM} %\textcolor{red}
{H.S. Dhar \emph{et al.}, New J. Phys. \textbf{15}, 013043 (2013).}

\bibitem{long-range} %\textcolor{red}
{X. Chen \emph{et al.}, Phys. Rev. B 82, 155138 (2010).}

\bibitem{DLC}%\textcolor{blue}
{J. Cui \emph{et al.}, Nat. Comm. \textbf{3}, 812 (2012); A. Hamma \emph{et al.}, Phys. Rev. Lett. \textbf{110}, 210602 (2013).}

\bibitem{bhaskar} G. Baskaran, Indian J. Phys. \textbf{89}, 583 (2006).

\bibitem{EP} J.-L. Song, 
%\emph{et al.},
S.-J. Gu, and H.-Q. Lin,
Phys. Rev. B \textbf{74}, 155119 (2006);
A. Tribedi and I. Bose, Phys. Rev. A \textbf{79}, 012331 (2009); 
A. B. Kallin, 
%\emph{et al.},
I. Gonzalez, M. B. Hastings, and R. G. Melko, 
Phys. Rev. Lett. \textbf{103}, 117203 (2009);
I. A. Kov{\`a}cs and F. Igl{\`o}i, Phys. Rev. B \textbf{80}, 214416 (2009);
D. Poilblanc, Phys. Rev. Lett. \textbf{105}, 077202 (2010); 
H. Katsura, 
%\emph{et al.},
N. Kawashima, A. N. Kirillov, V. E. Korepin, S. Tanaka,
J. Phys. A: Math. Theor. \textbf{43}, 255303  (2010);
%H. S. Dhar and A. Sen(De), \emph{ibid.} \textbf{44}, 465302 (2010);
A. M. L{\"a}uchli and J. Schliemann, arXiv:1106.3419,
and references therein.

\bibitem{GGM1} R. Prabhu, S. Pradhan, A. Sen(De), and U. Sen, Phys. Rev. A {\bf 84}, 042334 (2011); 
 %
R. Prabhu, A.K. Pati, A. Sen(De), and U. Sen, arXiv:1109.4318 (accepted in Phys. Rev. A); 
%
M. N. Bera, R. Prabhu, A. Sen(De), and U. Sen, Phys. Rev. A {\bf 86}, 012319 (2012);
R. Prabhu, A. Sen(De), and U. Sen, arXiv:1208.6535 (2012);

\bibitem{GGM2} H. S. Dhar and A. Sen(De), J. Phys. A: Math. Theor. {\bf 44}, 465302 (2011); M. N. Bera, R. Prabhu, A. Sen(De), and U. Sen, arXiv:1209.1523 (2012); A. Biswas, R. Prabhu, A. Sen(De), and U. Sen, arXiv:1211.3241 (2012).


\end{thebibliography}
\end{document}